\documentstyle[12pt]{article}
\catcode`@=11
\def\section{\setcounter{equation}{0}
\@startsection {section}{1}{\z@}{-3.5ex plus -1ex minus
 -.2ex}{2.3ex plus .2ex}{\large\bf}}
\newcommand{\be}{\begin{equation}}
\newcommand{\ee}{\end{equation}}
\newcommand{\bes}{\begin{eqnarray}}
\newcommand{\ees}{\end{eqnarray}}

\newcommand{\nn}{\nonumber }

\begin{document}
\title{
Thermodynamical Properties of the Antiferromagnetic Heisenberg Model in
Dimensions $d=
1,2,3$\\[2cm]
}
\author{M.Karbach\thanks{e-mail: karbach@wpts0.physik.uni-wuppertal.de}
, K.-H. M\"utter and P. Ueberholz\\Physics Department, University of
Wuppertal\\
D-5600 Wuppertal 1, Germany\\[1cm]
H.Kr\"oger\\Max Planck Institut f\"ur Kernphysik\\D-6900 Heidelberg,
Germany\\[2cm]}

\date{WU B 93-12, March 93}

\maketitle

{\em
{\centerline{\bf Abstract}
\vspace{1cm}}

The evolution equation for the expectation values of the Boltzmann factor
between valence bond
states is evaluated in lowest order of the dimer cluster expansion. Explicit
formulas are given
for the internal energy and the specific heat of the $d$-dimensional
antiferromagnetic
Heisenberg model.
}
\newpage
\section{Introduction}
A new approach to the spin $\frac{1}{2}$ quantum Heisenberg model
with Hamiltonian:
\be
H=\sum_{<x,y>} 4(x,y)
\ee
where:
\be
(x,y)=\frac{1}{4}(1-\vec{\sigma}(x)\vec{\sigma}(y))
\ee
has been proposed by one of us in the preceding paper [1], which is
refered as paper 1 in the following. In this second paper we are going to
exploit the new method in the sector with total spin $0$. It was pointed
out in paper 1, that the partition function in the spin $0$ sector:
\be
{\rm tr \ } (\exp \beta H_0)=\frac{1}{a_0}\sum_K f(\beta,K)
\ee
contains already the whole information on the thermodynamics of the
antiferromagnetic Heisenberg model in the absence of an external
field -- i.e. for vanishing magnetization $M=\frac{S}{V}=0$. Moreover,
it was shown that the partition function
(1.3)
is obtained from the expectation values of the Boltzmann factor
 $\exp(\beta H_0)$
between valence bond states $|K\rangle$:
\be
f(\beta,K)=\langle K|\exp(\beta H_0)|K\rangle.
\ee
These expectation values  were proven to obey an evolution equation :
\bes
\frac{\partial}{\partial \beta}f(\beta,K)
=&&(dV+3N_1(K))f(\beta,K) \nn \\ &&+
   \sum_{<x,y>}\left(f(\beta,Q_+(x,y)K)-f(\beta,Q_-(x,y)K)\right)
\ees
where $Q_{\pm}(x,y)$ are permutation operators which act on
the valence bond configuration $K$, as defined in eqs.(2.13,14) of paper 1.
$N_1(K)$ is the number of dimers. These are the nearest neighbour
valence bonds on the configuration $K$.\\
It is clear from the evolution equation (1.5) that the expectation values
$f(\beta,K)$ increase much stronger with $\beta$ on configurations with
a large dimer density $x=\frac{N_1}{V}$. On the other hand, these
configurations have low entropy, since the number
of valence bond configurations $\nu(N_1,V)$ with a given number of
dimers $N_1$ decrease as $(N_1!)^{-1}$. Therefore, if we assume that
the averages:
\be
f(\beta,N_1)=\frac{1}{\nu(N_1,V)}\sum_{K(N_1)} f(\beta,K(N_1))
\ee
over all valence bond configurations $K(N_1)$
 with  a fixed number
$N_1$
of dimers behave in the
combined limit:
\be
V,N_1,n \rightarrow \infty , \quad x=N_1/V,
\quad \beta  \quad {\rm fixed}
\ee
as:
\be
f(\beta,N_1)=\exp(V\phi(\beta,x))
\ee
one is led to the conclusion that
the zero density $(x=0)$
contribution of the dimers
determines the thermodynamical properties of the model. This statement
holds in general provided that the function
$\phi(\beta,x)$ is differentiable with respect to $x$.\\
E.g. the internal energy per site at fixed inverse temperature $\beta$
is given by:
\be
U(\beta)=\frac{\partial}{\partial \beta} \phi(\beta,x=0).
\ee
We are not yet in the position to evaluate the exact evolution equation
(1.5) in the thermodynamical limit. However it will be shown in section
2 that the lowest order in the `dimer cluster expansion' can be solved
analytically in the thermodynamical limit. Results for the ground-state
energy and the specific heat of the AFH-model in dimensions $d=1,2,3$
are presented in sections 3 and 4.
\section{The Dimer Cluster Expansion in the Spin $0$ Sector}
In paper 1 an approximation scheme was proposed for the evaluation
of the evolution equation (1.5) in the thermodynamical limit. In lowest
order of this `dimer cluster expansion' (1.5)
is averaged over all valence bond configurations $K$ with a fixed
number $N_1(K)$ of dimers. The approximation would be exact if the
right hand side of (1.5) could be expressed as well in terms of
the averages (1.6). This is not possible for all the terms in
the sum on the right hand side of (1.5). Here we approximate
the $f(\beta,K)$ by their average (1.6). The approximate evolution
equation can be brought into the form:
\be
\frac{\partial}{\partial \beta }f(\beta,N_1)=(dV+3N_1)f(\beta,N_1)+
V\sum_j c_j(N_1,V) f(\beta,N_1+j).
\ee
In the combined limit (1.7) the coefficients $c_j(N_1,V)$ were found to be\\
(eqs(5.5-11) of paper 1):
\bes
c_{-2}(x)=&&-w_1, \\
c_{-1}(x)=&&3w_1-w_2-2(2d-1)x  ,    \\
c_0(x)=&&-3w_1+w_2+4(2d-1)x-d+x  , \\
c_1(x)=&&w_1-2(2d-1)x+d-x .
\ees
Here $w_1(x)$ and $w_2(x)$ are the probabilities to find
on the valence bond configurations $K(N_1)$
dimer pairs with arbitrary and parallel orientation, respectively.
The approximate evolution equation (2.1) can be solved analytically in the
combined limit (1.7) if we assume that the averages (1.6) have the form
(1.8). Then eq.(2.1) yields a partial differential equation for
$\phi(\beta,x)$:
\be
\frac{\partial}{\partial x}\phi(\beta,x)=\log R
\ee
\be
\frac{\partial}{\partial \beta}\phi(\beta,x)= L(x,R)
\ee
where
\be
L(x,R)=d+3x + \sum_j c_j(x)R^j
\ee
with the initial condition:
\be
\phi(\beta=0,x)=0.
\ee
The solution proceeds as follows.
We first introduce $R$ instead of $\beta=\beta(x,R)$ as a new variable by
means of the Legendre transform:
\be
\Phi(x,R)=\phi(\beta,x)-\beta \frac{\partial}{\partial \beta}\phi(\beta,x).
\ee
Then we find for the partial derivatives:
\be
\left.\frac{\partial\Phi}{\partial x}\right|_R=\log R-\beta
\frac{\partial}{\partial x}
   L(x,R),
\ee
\be
\left.\frac{\partial\Phi}{\partial R}\right|_x =-\beta
\frac{\partial}{\partial R} L(x,R).
\ee
If we eliminate $\beta$:
\be
\frac{\partial \Phi}{\partial R} \frac{\partial L}{\partial x}
=\left(\frac{\partial \Phi}{\partial x} -\log R\right)
\frac{\partial L}{\partial R}.
\ee
we get a partial differential equation for $\Phi(x,R)$ with initial
condition:
\be
\Phi(x,R=1)=0.
\ee
A solution of eqs.(2.13), (2.14) is easily found:
\be
\Phi(x,R)=x\log R- \int\limits_{1}^{R} \frac{dR'}{R'}g(R',L(x,R))
\ee
provided that $g(R',L(x,R))$ solves the implicit equation:
\be
x=g(R'=R,L(x,R)).
\ee
Therefore we get the integrand $g(R,L)$ in eq.(2.15) from
\be
L(g,R)=L.
\ee

\section{Thermodynamical Properties in One Dimension}
The probability $w_1(x)$ to find dimer pairs on the valence bond
configurations $K(N_1)$ can be calculated analytically in the case
$d=1$, as is
done in appendix A:
\be
w_1(x)=\frac{x^2}{1-x}.
\ee
The implicit equation (2.17) for $g(R,L)$ turns out to be quadratic
and can be easily solved:
\be
g_{\pm}(R,L)=1-\frac{1}{2a(R)}\left(L-b(R) \pm \sqrt{D(R,L)}\right)
\ee
where
\be
a(R)=4R-11+5R^{-1}-R^{-2},
\ee
\be
b(R)=-4R+14-8R^{-1}+2R^{-2},
\ee
\be
D(R,L)=(L-b(R))^2-4a(R)\frac{(R-1)^3}{R^2}.
\ee
The solution has to satisfy eq.(2.16) which means in particular for:
\be
x=0, \quad L(x=0,R)=R, \quad g(R,L=R)=0.
\ee
We have plotted in Fig.1 both solutions given by eq.(3.2). The condition
(3.6) is satisfied for:
\be
g_+(R,L=R)=0 \quad {\rm if} \quad 1<L<L_1=2.387425
\ee
\be
g_-(R,L=R)=0 \quad {\rm if} \quad L_1<L<L_2=2.588229.
\ee
Both solutions meet each other at $R=R_2(L)$, which defines the zero
of $D(R,L)$. This zero produces a branch point singularity of
$g(R,L)$ in the complex $R$-plane, which has to be taken into account
if we define the path of integration for the solution (2.15):
\be
\Phi(x,R)=x\log R- \int\limits_{1}^{R} \frac{dR'}{R'}g_+(R',L)\quad
  {\rm for} \quad L= L(x,R)<L_1
\ee
\bes
\Phi(x,R)=&&x\log R- \int\limits_{1}^{R_2(L)} \frac{dR'}{R'}g_+(R',L)
-\int\limits_{R_2(L)}^{R} \frac{dR'}{R'}g_-(R',L) \\ &&
\quad {\rm for} \quad L_1<L(x,R)<L_2 \nn
\ees
Finally, we need the relation (2.12) for going back to the original
variable, the inverse temperature $\beta$:
\be
\beta(x,R)=\int\limits_{1}^{R} \frac{dR'}{R'}\frac{\partial}{\partial
L}g_+(R',L)
\quad {\rm for} \quad L=L(x,R)<L_1
\ee
\bes
\beta(x,R)=&& \int\limits_{1}^{R_2(L)} \frac{dR'}{R'}\frac{\partial}{\partial
L}
g_+(R',L)
+\int\limits_{R_2(L)}^{R} \frac{dR'}{R'} \frac{\partial}{\partial L}g_-(R',L)
\\
&&\quad {\rm for} \quad L_1<L(x,R)<L_2 .\nn
\ees
Since $R$ and $L$ vary only in open and finite intervals:
\be
1<R<R_2(L_2)=3.772126 ,\quad  1<L<L_2
\ee
the derivatives (2.6) and (2.7) are finite and the premise for eq.(1.9)
is satisfied. Therefore the thermodynamical properties are obtained
from the solution $\phi(\beta,x)$ at vanishing dimer density:
\be
x=0,\quad  L= R=\frac{\partial }{\partial \beta} \phi(\beta,x=0)=U(\beta).
\ee
The inverse temperature $\beta (U)$ as function of the internal energy
$U$ can be taken from eqs.(3.11,12) for
($x=0,L=R=U$). This function is shown in Fig.2. It develops a singularity
for
\be
R=U \rightarrow L_2, \quad \beta \rightarrow \infty.
\ee
which can be identified with the low temperature
limit. Moreover we see from eq.(1.9) that $L_2 $ is  just the
ground state energy per site $h_0=L_2=2.588229$. This value is $7\%$
below the exact value $h_0=4\log 2$ which gives us an idea on the accuracy
of the lowest order in the dimer cluster expansion. In Fig.3 we present
the specific heat as function of the temperature\footnote{Our
definition (1.1) and (1.3) of $H$ and $\beta$ differ from the usual one.
Our $\beta$ is related to the  usual definition of the temperature via:
$\beta=(2dT)^{-1}$.}. Position and height
of the maximum are:
\be
T=0.982 (0.962) \quad C=0.296 (0.350)
\ee
For comparison we have listed in brackets the estimates of ref. [2],
obtained from small rings up to $16$ sites. The low temperature behavior
[3]
of the specific heat, as it is predicted by conformal invariance
and derived rigorously by J. Suzuki, Y. Akutsu and M. Wadati:
\be
C(T)=\frac{1}{3} T \quad {\rm for} \quad T\rightarrow 0
\ee
is represented in Fig.3 by the dotted line. The high temperature expansion
has been computed to order 21 [4] and is represented in Fig.3 by the
dashed line.
\section{Thermodynamical Properties in Two and Three Dimensions}
The probabilities $w_1(x),w_2(x)$ to find on the valence bond
configurations $K(N_1)$ dimer pairs with arbitrary and parallel
orientation are determined in appendix A:
\be
w_1(x)=(2d-1)^2\frac{x^2}{d-x}
\ee
\be
w_2(x)=2(d-1)\frac{x^2}{d-2x}.
\ee
The implicit equation (2.17)-together with (2.8) and (2.2-5)- turns out
to be of third order now:
\be
g^3a(R,d)+g^2(b(R,d)-2\frac{L}{d^2})
+g(c(R,d)+3\frac{L}{d})=L-dR
\ee
where
\be
a(R,d)=2\frac{(2d-1)^2}{R^2d^2}+\frac{1}{Rd^2}
(-24d^2+18d-4)
+\frac{1}{d^2}(24d^2-10d+8)-8R
\ee
\be
b(R,d)=-\frac{(2d-1)^2}{R^2d}+\frac{1}{dR}(12d^2-2d-1)
 -\frac{1}{d}(12d^2+10d+5)+4R(d+2)
\ee
\be
c(R,d)=\frac{2}{R}(1-2d)+8d-R(4d+2).
\ee
The real solutions $g_{\pm}(R,L)$ are shown in Figs. 4 and 5 for $d=2$
and $d=3$, respectively. We observe that the condition (2.16) for
$x=0,L(x=0,R)=dR$:
\be
g_{\sigma}(R,L=dR)=0,\quad \sigma=+,-
\ee
is satisfied for the solution:
\be
\sigma=+:\quad {\rm if} \quad 1<L<L_1(d)
\ee
\be
\sigma=-:\quad {\rm if} \quad L_1(d)<L<L_2(d)
\ee
where
\be
L_1(d=2)=3.625833,\quad L_2(d=2)=4.154262
\ee
\be
L_1(d=3)=4.862986,\quad L_2(d=3)=5.399962.
\ee
Both solutions meet each other at $R=R_2(L)$, where the derivative
$\frac{\partial g}{\partial R}$ divergies. At this point $g(R,L)$
has a branch point singularity in the complex $R$-plane which we
pick up in the path of integration for the solution (2.15). The
thermodynamical properties of the AFH model in d dimensions
are obtained from:
\be
\beta(x=0,R)=\int\limits_{1}^{R}\frac{dR'}{R'}\frac{\partial}{\partial
L}g_+(R',L)
 \quad {\rm for} \quad L=dR<L_1
\ee
\bes
\beta(x=0,R)=&&\int\limits_{1}^{R_2(dR)}\frac{dR'}{R'}\frac{\partial}{\partial
L}
g_+(R',L) \\  &&+ \int\limits_{R_2(dR)}^{R}\frac{dR'}{R'}
\frac{\partial}{\partial L}g_-(R',L)\quad {\rm for} \quad L_1<dR<L_2 . \nn
\ees
This function relates the inverse temperature $\beta $ with the
internal energy per site:
\be
 U(\beta)=\frac{\partial}{\partial \beta}\phi(\beta,x=0)=L=dR
\ee
which is shown in Fig.6  for $d=2$ and $d=3$.
A singularity appears for:
\be
dR\rightarrow L_2,\quad \beta(0,R)\rightarrow \infty
\ee
i.e. in the zero temperature
limit. $L_2$ is the groundstate energy per site in the lowest order
of the dimer cluster expansion (cf.(2.1)). The values listed in
eqs. (4.10,11) have to be compared with the estimates of ref.[5,6,7]
for the groundstate energy per site $e(d)$:
\be
h_0(d=2)=2-4e(d=2)=4.678,  \quad h_0(d=3)=3-4e(d=3)=6.611
\ee
The specific heats for $d=2$ and $d=3$ are shown in Figs.7 and 8,
respectively.\\
For $d=2$ the low temperature behavior is known from chiral perturbation
theory [8]:
\be
C(T)=\frac{6\zeta(3)}{\pi c^2}T^2 \quad {\rm for} \quad T\rightarrow 0
\ee
where
\be
c=1.68
\ee
is the spin wave velocity [5].\\
The high temperature expansion has been computed up to order $10$ for
$d=2$ [9] and $d=3$ [10]. The
high and low temperature limits are represented in Fig.7 by the dashed
and dotted curves. We observe good agreement in the high temperature
regime, whereas the low temperature behavior (4.17) cannot be reproduced
by the lowest order of the dimer cluster expansion.
\section{Summary and Perspectives}
In this paper we have started a first attempt to evaluate the evolution
equation (1.5) for the expectation values (1.4) of the Boltzmann factor
between valence bond states. To lowest order in the dimer cluster
expansion the thermodynamical limit of the evolution equation (2.1)
leads to a partial differential equation (2.6-7) which can be solved
analytically up to integrations. As a result, we obtain an explicit
representation (3.11-12, 4.12-13) of the inverse temperature $\beta$ as
function of the internal energy per site $U$ for the antiferromagnetic
Heisenberg model in dimensions $d=1,2,3$. The predictions of the lowest
order in the dimer cluster expansion for the specific heat are in
agreement with the high temperature expansion but they are in disagreement
with the low temperature expansion (3.17) and (4.17). This means, that
the formation of dimer clusters-which has not been taken into account
in the lowest order approximation (2.1)- plays an important role for
the correct description of the low temperature behavior. The specific
heat has a maximum at temperatures $T(d)$ which decreases with the
dimension $d$. The height of the maxima is almost independent of $d$,
the width shrinks with $d$. Therefore, we do not see any indications for a
phasetransition in our results for the specific heat in three dimensions,
as it was suggested from the hightemperature expansion of ref.[10].

\newpage
\begin{appendix}
\appendix{\large Appendix}
\section{Dimer Pairs on Valence Bond Configurations}
In this appendix we will compute the probability:
\be
w([1,2][3,4])=\nu(N_1)^{-1}\sum_{K(N_1)}
\delta([1,2][3,4]\in K(N_1))
\ee
to find a dimer pair at the neighbouring sites $2$ and $3$ on a
valence bond configuration $K(N_1)$ with $N_1$ dimers. $\nu(N_1)$
denotes the total number of these configurations.\\
Let us keep the $N_1$ dimers fixed in a `dimer configuration' $D(N_1)$
and ask how many possibilities we have to connect the remainig
$V-2N_1$ sites on the lattice by valence bonds.  This number is
approximately given by $(V-2N_1-1)!!$ i.e. by the total number of
valence bond configurations on the remaining sites. In a strict sense
we should count only those configurations on the remaining sites which
do not contain any further dimers. Indeed these configurations dominate
the thermodynamical limit, since configurations with dimers have lower
entropy. This argument tells us that the dimer probability (A.1)
on the valence bond configuration $K(N_1)$ is just the same as on
the dimer configuration $D(N_1)$:
\be
w([1,2][3,4])=\rho(N_1,V)^{-1}\sum_{D(N_1)}
\delta([1,2],[3,4]\in D(N_1))
\ee
where:
\be
\rho(N_1,V)=\sum_{D(N_1)}
\ee
is the number of possibilities to distribute $N_1$ dimers on a
lattice with $V$ sites. \\
For $d=1$ and open boundary conditions one finds:
\be
\rho_o(N_1,V)=\frac{(V/2+n_1)!}{(2n_1)!(V/2-n_1)!}
\ee
where $n_1=V-2N_1$ is the number of sites which are not occupied
with dimers.\\
For $d=1$ and periodic boundary conditions one finds:
\be
\rho(N_1,V)=\rho_o(N_1,V)+\rho_o(N_1-1,V-2)-\rho_o(N_1+1,V-2).
\ee
The second and third term take into account the different interpretation
of the valence bond $[1,V]$. It is a dimer for periodic but not for
open boundary conditions.\\
In the case $d=1$ the probability (A.3) to find a dimer pair turns
out to be:
\be
w([1,2][3,4])=\frac{\rho_o(N_1-2,V-4)}{\rho(N_1,V)}
\ee
which leads in the combined limit (1.7) to eq.(3.1).\\
In the case $d=2$ and $d=3$ the dimer pair probability (A.2) can be
determined in a simulation of the monomer dimer system. We have done
that on the lattices
\footnote{
The simulations on the $200^3$ lattice has been
performed on the connection Machine CM5. The results are based upon a test
version of the
software where the emphasis was on providing functionality. This software
release has not
had the benefit of optimization or performance.}:
\begin{itemize}
\item $100 \times 100,\quad 200 \times 200$
\item $50 \times 50 \times 50,\quad  200 \times 200 \times 200$
\end{itemize}

The results for the dimer pair probabilities $w_l(x),l=1,2$ together
with the fits \\(4.1-2) are shown in Fig. 9  for $d=2$ and in
Fig. 10  for $d=3$. The behavior of $w_1(x)$ for
$x=N_1/V \rightarrow 0$ and $x\rightarrow 1/2$ and that of $w_2(x)$
for $x\rightarrow 0$ can be easily checked by combinatorical
considerations. Note that the fit (4.1) for the dimer pair probability
$w_1(x)$ is perfect for $d=2$ and $d=3$. We therefore believe, that
(4.1) represents indeed the exact formula for this quantity.
On the other hand, the fit (4.2) for the parallel dimer pair probability
$w_2(x)$ is not perfect. We therefore expect that the exact formula
for this quantity is more complicated than the fit (4.2).
\newpage
\end{appendix}
{\bf Figure Captions}
\begin{enumerate}
\item The real solutions (3.2) of the implicit equations (2.17) for
$d=1 $ . The solutions coincide on the solid curve $R=R_2(L)$
where the square root in eq. (3.2) vanishes.
\item The inverse temperature $\beta(U)$ versus the internal energy per
site $U$, as it follows from (3.11,12) for $x=0,U=L=R,
d=1$.
\item The specific heat of the one-dimensional AFH-model in lowest order
of the dimer cluster expansion. The dotted curve represents the low
temperature behavior (3.17), the dashed curve the high temperature
expansion of ref.[4].
\item The real solutions $g_{\pm}(R;L)$ of the implicit equation (4.3) for
$d=2$. The solutions coincide on the solid curve $R=R_2(L)$.
\item Same as Fig.4 for $d=3$.
\item The inverse temperature $\beta(U)$ versus the internal energy
per site $U$ as it follows from eqs. (4.12,13) for $x=0,U=L=dR$.
Solid curve: $d=2$, dotted curve: $d=3$.
\item The specific heat of the $d=2$ AFH model in lowest order of the
dimer cluster expansion. The dotted curve represents the low temperature
behavior (4.17); the dashed curve the high temperature expansion
of ref.[9].
\item Same as Fig.7 for $d=3$.
\item The dimer pair probabilities for $d=2$: solid curve: $w_1(x)$,
dotted curve: $w_2(x)$
\item Same as Fig.9 for $d=3$.
\end{enumerate}


\newpage
\begin{thebibliography}{99}
\bibitem{} K.-H. M\"utter, `A new approach to the quantum Heisenberg
model' University of Wuppertal preprint March 1993
\bibitem{} J.C. Bonner and M.E. Fisher, Phys. Rev. 135, A640 (1964);
K. Fabricius, U. L\"ow, K.-H. M\"utter and P. Ueberholz, Phys. Rev. B44,
7476 (1991); K. Fabricius, U.L\"ow and K.-H. M\"utter, Z.Phys.B to be
published
\bibitem{} I. Affleck, Phys, Rev. Lett. 56, 746 (1986);
 J.Suzuki, Y.Akutsu and M. Wadati, J. Phys. Soc. Jpn. 59,
2667 (1990); M. Takahashi, Prog. Theor. Phys. 50, 1519 (1973)
\bibitem{} G.A. Baker, G.S. Rushbrooke and H.E. Gilbert, Phys. Rev.
135, A1272 (1964)
\bibitem{} U.J. Wiese and H.P. Ying, `A determination of the low energy
parameters of the 2-d Heisenberg antiferromagnet' University of Bern
preprint Febr. 1993
\bibitem{} R.P.Singh, Phys. Rev. B39, 9760 (1989); Zheng Weihong,
J. Oitma and C.J. Hamer, Phys. Rev. B43, 8321 (1991);
H.J. Schulz and T.A.L. Ziman, Europhys. Lett. 18, 355 (1992);
O. Haan, J.U. Klaetke and K.-H. M\"utter, Phys. Rev. B46
5723 (1992)
\bibitem{} Zheng Weihong, J.Oitma and C.J. Hamer, Phys. Rev. B44, 11869
(1991)
\bibitem{} P.Hasenfratz and F. Niedermayer, `Finite size and temperature
effects in the AF Heisenberg model', University of Bern preprint
Febr. 1993; D.S. Fisher, Phys. Rev. B39, 11783 (1989)
\bibitem{} T. Barnes, Int. J. Mod. Phys. C2, 659 (1991)
\bibitem{} G.S. Rushbrooke,
G.A. Baker and P.J. Wood in `Phase transitions and critical phenomena'
Vol.3, eds. C. Domb and M.S. Green, Academic Press (1974).
\end{thebibliography}
\end{document}